\documentclass[10pt,journal]{IEEEtran}
\usepackage[ruled,linesnumbered]{algorithm2e}
\usepackage{amsthm}
\usepackage{amsmath}
\usepackage{graphicx}
\usepackage{epsfig,graphics,psfrag,amsmath,amssymb}
\usepackage{cite}
\usepackage{float}
\usepackage{tocvsec2}
\usepackage{color}
\usepackage{verbatim}
\usepackage{mathtools}
\usepackage{hhline}
\usepackage{enumerate}
\usepackage{arydshln}
\usepackage{bbm}
\usepackage{stackengine}
\usepackage{mathrsfs}
\usepackage{enumitem}
\usepackage{caption}
\usepackage{stfloats}
\usepackage{amssymb}
\usepackage{hyperref}       
\usepackage[export]{adjustbox} 

\usepackage{setspace}

\usepackage[symbol]{footmisc}
\renewcommand{\thefootnote}{\fnsymbol{footnote}}
\newcommand \footnoteONLYtext[1]
{
	\let \mybackup \thefootnote
	\let \thefootnote \relax
	\footnotetext{#1}
	\let \thefootnote \mybackup
	\let \mybackup \imareallyundefinedcommand
}

\makeatletter

\newcommand{\Rmnum}[1]{\expandafter\@slowromancap\romannumeral #1@}
\makeatother

\makeatletter
\newcommand*\bigcdot{\mathpalette\bigcdot@{.5}}
\newcommand*\bigcdot@[2]{\mathbin{\vcenter{\hbox{\scalebox{#2}{$\m@th#1\bullet$}}}}}
\makeatother

\def\@#1{\pmb{#1}}
\def\b#1{\mathbb{#1}}
\def\bf#1{\mathbf{#1}}
\def\s#1{\mathsf{#1}}

\usepackage{amsmath,graphicx}
\usepackage{amsmath,amssymb,amsfonts}
\usepackage{cite, soul}
\usepackage{url}
\usepackage{amsthm}
\usepackage{bm}
\usepackage{comment}
\usepackage{multirow}
\usepackage{booktabs}
\usepackage{hhline}
\usepackage{graphicx}     
\usepackage{caption}      
\usepackage{subcaption}   

\usepackage{setspace}
\usepackage{setspace}
\usepackage[ruled,linesnumbered]{algorithm2e}

\def\@#1{\pmb{#1}}
\def\b#1{\mathbb{#1}}
\def\bf#1{\mathbf{#1}}
\def\s#1{\mathsf{#1}}

\title{Multimodal Mixture-of-Experts for ISAC in Low-Altitude Wireless Networks}
\author{Kai Zhang, \emph{Graduate Student Member}, \emph{IEEE}, Wentao Yu, \emph{Member}, \emph{IEEE}, Hengtao He, \emph{Member}, \emph{IEEE},\\ Shenghui Song, \emph{Senior Member}, \emph{IEEE}, Jun Zhang, \emph{Fellow}, \emph{IEEE}, and Khaled B. Letaief, \emph{Fellow}, \emph{IEEE}
}

\begin{document}


\maketitle

\footnoteONLYtext{
	\hspace*{-1.85em}
	
	This work was presented in part at IEEE International Symposium on Personal, Indoor, Mobile Radio Communications (PIMRC), Istanbul, Turkey, Sept. 2025 \cite{kaiicc}.

	Kai Zhang, Shenghui Song, Jun Zhang, and Khaled B. Letaief are with the Department of Electronic and Computer Engineering, The Hong Kong University of Science and Technology, Clear Water Bay, Hong Kong (email: kzhangbn@connect.ust.hk, eeshsong@ust.hk, eejzhang@ust.hk, eekhaled@ust.hk).
	
	Wentao Yu was with the Department of Electronic and Computer Engineering, The Hong Kong University of Science and Technology, Clear Water Bay, Hong Kong. He is now with the Department of Electrical and Computer Engineering, The University of British Columbia, Vancouver, BC V6T 1Z4, Canada (e-mail: wentaoyu@ece.ubc.ca).
	
	Hengtao He is with the National Mobile Communications Research Laboratory, Southeast University, Nanjing, China (hehengtao@seu.edu.cn).
}

%
%
%
%
%

\begin{abstract}
	Integrated sensing and communication (ISAC) is a key enabler for low-altitude wireless networks (LAWNs), providing simultaneous environmental perception and data transmission in complex aerial scenarios.
	By combining heterogeneous sensing modalities such as visual, radar, lidar, and positional information, multimodal ISAC can improve both situational awareness and robustness of LAWNs. 
	However, most existing multimodal fusion approaches use static fusion strategies that treat all modalities equally and cannot adapt to channel heterogeneity or time-varying modality reliability in dynamic low-altitude environments.
	To address this fundamental limitation, we propose a mixture-of-experts (MoE) framework for multimodal ISAC in LAWNs.
	Each modality is processed by a dedicated expert network, and a lightweight gating module adaptively assigns fusion weights according to the instantaneous informativeness and reliability of each modality.
	To improve scalability under the stringent energy constraints of aerial platforms, we further develop a sparse MoE variant that selectively activates only a subset of experts, thereby reducing computation overhead while preserving the benefits of adaptive fusion.
	Comprehensive simulations on three typical ISAC tasks in LAWNs demonstrate that the proposed frameworks consistently outperform conventional multimodal fusion baselines in terms of learning performance and training sample efficiency.
\end{abstract}


\section{Introduction}

The rapid advancement of sixth-generation (6G) wireless networks is poised to revolutionize communication systems, shifting from conventional two-dimensional terrestrial networks to three-dimensional ubiquitous connectivity \cite{letaief2019roadmap,letaief2021edge}.
Within this landscape, low-altitude wireless networks (LAWNs) are emerging as a critical infrastructure.
By deploying unmanned aerial vehicles (UAVs) as flexible wireless relays, LAWNs form resilient three-dimensional communication links that bridge the coverage gaps of fixed terrestrial infrastructure \cite{8869712,yuan2025groundskyarchitecturesapplications}.
This architecture enables rapid deployment of wireless coverage in disaster-affected and remote areas where fixed infrastructure is unavailable or damaged \cite{8669870}.
Consequently, LAWNs are increasingly used to support applications such as intelligent aerial logistics \cite{kuru2019analysis}, precision agriculture \cite{zhang2012application}, and large-scale infrastructure monitoring \cite{zhang2020drone,9598833}.

Despite great potentials, LAWNs must maintain reliable wireless links while simultaneously performing real-time environmental perception in congested low-altitude airspace.
To tackle this challenge, integrated sensing and communication (ISAC) has emerged as a new paradigm that promotes mutual benefits between wireless communication and sensing within a unified system \cite{9737357,9540344}.
This is particularly important for LAWNs, where the complex aerial conditions render isolated communication or sensing unreliable.
Specifically, the sensed spatial information can stabilize wireless links in high-mobility scenarios by enabling proactive beam alignment and blockage prediction.
In addition, reliable connectivity enables real-time fusion of distributed sensing data, extending the sensing coverage of individual UAVs.

Given these promising capabilities, extensive research has investigated the integration of ISAC into LAWN systems, including advanced beamforming techniques \cite{9916163}, UAV trajectory optimization \cite{9858656}, and spectrum-efficient resource allocation \cite{10453349}.
Specifically, a substantial body of works have investigated sensing-aided communication, utilizing environmental awareness to facilitate robust UAV connectivity in high-mobility aerial scenarios \cite{10806828,10476610, jiang2025lowaltitudeuavtrackingsensingassisted, 10226306}.
For instance, by leveraging historical sensing data and extended Kalman filtering (EKF) to predict target motion, Zhou \emph{et al.} \cite{10806828} enabled precise beamforming without pilots to boost communication throughput, while Jiang \emph{et al.} \cite{10476610} optimized UAV trajectory to guarantee link maintenance.
Focusing on reliability, Jiang \emph{et al.} \cite{jiang2025lowaltitudeuavtrackingsensingassisted} maximized communication outage capacity by optimizing the sensing duration and UAV trajectory based on the outage probabilities predicted from sensing data.
Furthermore, Meng \emph{et al.} \cite{10226306} proposed a two-stage sensing protocol to acquire target location information, thereby optimizing the phase shifts of intelligent surface to maximize the achievable communication rate.
Conversely, other studies explore communication-aided sensing, exploiting ubiquitous wireless infrastructure to extend environmental perception coverage and precision \cite{ji2025dopplerbasedmultistaticdronetracking,11203895,10615952}.
For instance, Ji \emph{et al.} \cite{ji2025dopplerbasedmultistaticdronetracking} utilized existing long-term evolution (LTE) communication signals to track drone trajectories by analyzing Doppler frequency shifts, demonstrating that communication waveforms can be repurposed for high-precision sensing without dedicated hardware.
Extending this to networked cooperation, Yang \emph{et al.} \cite{11203895} proposed fusing signals from distributed communication nodes to improve sensing accuracy.
To further exploit the potential of infrastructure cooperation, Lu \emph{et al.} \cite{10615952} developed a symbol-level fusion method that utilizes the distributed network to achieve superior UAV localization and velocity estimation compared to single-station approaches.
While extensive studies have advanced the development of ISAC in LAWN systems, they primarily focused on unimodal sensing that leverages only radio frequency (RF) signals, extracting environmental features solely from communication signals such as channel state information (CSI), received signal strength, or Doppler shifts.
Although RF-based sensing is ubiquitous, it provides only coarse environmental perception and is sensitive to blockage, multipath scattering, and low signal-to-noise ratios (SNRs).
Consequently, reliance on a single sensing modality leads to performance degradation under high-speed aerial dynamics, compromising both the wireless connectivity and perception reliability essential for LAWNs.

To mitigate these limitations, multimodal ISAC integrates complementary sensing modalities to enhance both environmental perception and communication reliability in LAWNs \cite{10330577,11133431}.
By synergizing heterogeneous sensing sources (e.g., RF signals, visual images, and lidar point clouds), this approach exploits the unique strengths of different modalities to construct a high-fidelity spatial representation.
For instance, the weather resilience of radar can effectively offset visual degradation in adverse conditions, while the high angular resolution of lidar resolves the spatial ambiguity typical of RF signals.
Such fusion improves situational awareness and robustness to modality-specific failures that occur in dynamic aerial scenarios.
Several recent works have investigated the design of multimodal ISAC frameworks, developing different fusion architectures to align and integrate heterogeneous sensing modalities \cite{10912462,10577431,10539181,10829757,11208675}.
For instance, Zhu \emph{et al.} \cite{10912462} proposed a cross-modal feature enhancement mechanism that fuses image and radar data for multimodal beam prediction tasks, demonstrating improved robustness under varying modality quality.
Moving towards more unified representations, Tariq \emph{et al.} \cite{10577431} and Cui \emph{et al.} \cite{10539181} explored transformer-based architectures that fuse multiple sensing modalities for accurate beam prediction and reliable communication in ISAC systems.
To further address the generalization limitations of task-specific models, Cheng \emph{et al.} \cite{10829757} introduced multimodal large language models (MLLMs) into ISAC, leveraging their scalable cross-modal capabilities.
Specific to the low-altitude airspace, Xie \emph{et al.} \cite{11208675} developed a transformer-based perception fusion network that aligns vision, lidar, and radar data for proactive beamforming in dynamic LAWN scenarios.

However, most existing multimodal fusion methods employ static fusion strategies that treat all modalities equally and cannot adapt to rapid environmental changes or time-varying modality reliability in low-altitude environments.
To address this limitation, we propose a multimodal mixture-of-experts (MoE) architecture that enables adaptive cross-modal fusion for ISAC in LAWNs.
Unlike static baselines, the proposed architecture orchestrates a set of modality-specific expert networks through an adaptive gating mechanism that assigns data-dependent fusion weights to each modality.
This design allows the system to assess the instantaneous informativeness of each modality and place greater weight on reliable signals while down-weighting degraded ones. As a result, the fused representation becomes more robust to large fluctuations in sensing quality on high-speed aerial platforms in LAWNs.

\subsection{Contributions}

The main contributions of this paper are summarized as follows.

1. We establish a unified multimodal learning framework for ISAC in LAWNs by modeling the relationship between heterogeneous sensing modalities and RF communication signals as a generalized cross-modal prediction problem.
This formulation provides a common optimization paradigm that seamlessly supports diverse perception and communication tasks in low-altitude ISAC systems, including sensing-aided beam prediction, channel estimation, and communication-aided UAV trajectory tracking. 

2. We propose a novel multimodal MoE architecture that orchestrates modality-specific expert networks through an adaptive gating mechanism, enabling dynamic context-aware fusion based on the instantaneous informativeness and reliability of each modality. 
To meet the energy and computation constraints of aerial platforms, we further develop a sparse MoE variant that activates only a subset of experts during inference while preserving the benefits of adaptive fusion.

3. We conduct extensive simulations on representative ISAC tasks in realistic LAWN scenarios, including sensing‑aided beam prediction, sensing‑aided path loss prediction, and communication‑aided UAV trajectory tracking. Simulation results demonstrate that the proposed multimodal MoE and sparse MoE frameworks consistently outperform conventional multimodal fusion baselines in terms of learning performance and training sample efficiency.

\subsection{Organization and Notations}
The remainder of this paper is organized as follows.
Section II elaborates on the system model and formulates the unified learning framework for ISAC in LAWNs. Section III presents the proposed multimodal MoE architecture. Section IV introduces the sparse MoE variant to address the resource constraints of aerial platforms. Simulation results are presented in Section V, and we conclude the paper in Section VI.

\emph{Notations:} Column vectors and matrices are denoted by boldface lowercase and boldface capital letters, respectively.
The symbol $\b{R}$ denotes the set of real numbers.
$\b{C}^{M\times N}$ represents the space of the $M \times N$ complex-valued matrices.
$(\cdot)^\s{T}$ and $(\cdot)^\s{H}$ stand for the transpose and the conjugate transpose of their arguments, respectively. $\textup{Tr}(\bf{A})$ denotes the trace of matrix $\bf{A}$.
$\b{E}[\cdot]$ denotes the expectation operation. $\nabla$ represents the gradient operator. $|\cdot|$ and $\|\cdot\|$ stand for the $\ell_1$ and $\ell_2$ norm of vectors, respectively.


\section{System Model and Problem Formulation}\label{sec_2}

In this section, we first elaborate on the multimodal ISAC system model in LAWNs. We then establish a unified multimodal learning framework that incorporates heterogeneous sensing modalities and RF communication signals to support diverse perception and communication tasks.
Finally, we instantiate this framework through three representative ISAC tasks to demonstrate its broad applicability.

\subsection{Multimodal ISAC System Model}

\begin{figure*}[t]
	\renewcommand\figurename{Fig.}
	\centering \vspace*{5pt} \setlength{\baselineskip}{10pt}
	\includegraphics[width = 1\textwidth,trim=15 5 37 10,clip]{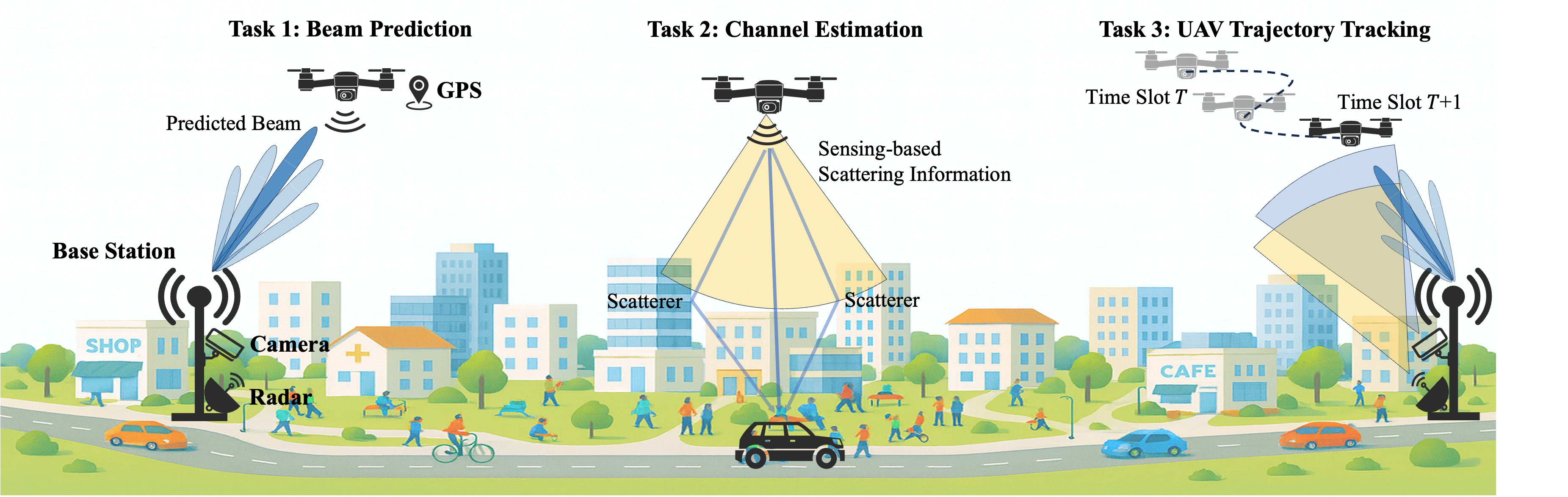}
	\caption{An illustration of multimodal ISAC in LAWNs, where a ground BS and UAVs exploit heterogeneous sensing data to support sensing-aided beam prediction, sensing-aided channel estimation, and communication-aided UAV trajectory tracking.}\label{fig_illustration}
	\vspace{6pt}
\end{figure*}

We consider a multimodal ISAC system deployed in LAWNs, as shown in Fig.~\ref{fig_illustration}.
The system comprises a ground base station (BS) equipped with $M$ antennas and $K$ single-antenna authorized UAVs, denoted by the set $\mathcal{K} = \{1,\dots,K\}$.
The BS conducts downlink communication with the UAVs while simultaneously acquiring spatial and geometric information of the propagation environment.
In addition to RF transceivers, both the BS and UAVs are equipped with heterogeneous sensors, such as cameras, lidars, radars, and global positioning systems (GPS).
These heterogeneous sensing platforms collectively provide multimodal observations of the surrounding environment, which are exploited to enhance both environmental perception and communication performance. The communication and sensing models are detailed as follows.

\subsubsection{Communication Model}

We consider a frequency-division duplexing millimeter-wave (mmWave) communication system operating over a flat block-fading channel.
For the downlink data transmission, let $\mathbf{V}(t) = [\mathbf{v}_1(t), \dots, \mathbf{v}_K(t)] \in \mathbb{C}^{M\times K}$ denote the beamforming matrix at the BS in time slot $t$, where each column vector $\mathbf{v}_k(t) \in \mathbb{C}^{M\times 1}$ corresponds to the beamforming vector designated for the $k$-th UAV.
Then, the transmitted signal vector is given by
\begin{equation}
	\mathbf{x}(t) = \sum_{k=1}^{K}\mathbf{v}_k(t)s_k(t) = \mathbf{V}(t)\mathbf{s}(t),
\end{equation}
where $\mathbf{s}(t) = [s_1(t), \dots, s_K(t)]^\s{T} \in \mathbb{C}^{K\times 1}$ is the vector of transmitted symbols with $\mathbb{E}[|s_k(t)|^2] = 1$. The BS transmission power is constrained by
\begin{equation}
	\textup{Tr}\big(\mathbf{V}(t)\mathbf{V}^{\s{H}}(t)\big) \le P_{\text{DL}},
\end{equation}
where $P_{\text{DL}}$ denotes the maximum downlink transmission power.

Let $\mathbf{h}_k(t) \in \b{C}^{M\times 1}$ denote the downlink channel between the BS and UAV $k$ at time slot $t$.
We adopt a geometric multi-path channel model \cite{9048929}, under which the downlink channel is expressed as
\begin{equation}
	\mathbf{h}_k(t)
	= \sum_{\ell=1}^{L_p} \alpha_{\ell,k}(t)\, e^{-j 2\pi f_c \tau_{\ell,k}(t)} \mathbf{a}\big(\theta_{\ell,k}(t)\big),
\end{equation}
where $L_p$ is the number of propagation paths, $f_c$ is the carrier frequency, $\alpha_{\ell,k}(t)$ is the path loss of the $\ell$-th path, $\tau_{\ell,k}(t) = d_{\ell,k}(t)/c$ is the associated propagation delay with $d_{\ell,k}(t)$ denoting the path length and $c$ the speed of light, and $\theta_{\ell,k}(t)$ is the angle of departure (AoD). For a uniform linear array (ULA) with $M$ antenna elements and inter-element spacing $d_{\textup{ant}}$, the array response vector corresponding to AoD $\theta_{\ell,k}(t)$ is given by
\begin{equation}
	\begin{aligned}
		&\mathbf{a}\big(\theta_{\ell,k}(t)\big)\\
		&= \big[1, e^{j\frac{2\pi}{\lambda} d_{\textup{ant}} \sin(\theta_{\ell,k}(t))}, \dots,
		e^{j\frac{2\pi}{\lambda} d_{\textup{ant}} (M-1) \sin(\theta_{\ell,k}(t))} \big]^{\s{T}},
	\end{aligned}
\end{equation}
where $\lambda$ is the carrier wavelength.
Accordingly, the received signal at UAV $k$ at time slot $t$ can be written as
\begin{equation}
	y_k(t) = \mathbf{h}_k^{\s{H}}(t)\mathbf{v}_k(t)s_k(t)
	+ \sum_{j\neq k} \mathbf{h}_k^{\s{H}}(t)\mathbf{v}_j(t)s_j(t)
	+ z_k(t),
\end{equation}
where $z_k(t)\sim \mathcal{CN}(0,\sigma_0^2)$ denotes the additive white Gaussian noise with variance $\sigma_0^2$. Consequently, the instantaneous downlink sum-rate is expressed as
\begin{equation}
	R(t) = \sum_{k=1}^{K}
	\log_2\!\left(
	1+\frac{\big|\mathbf{h}_k^{\s{H}}(t)\mathbf{v}_k(t)\big|^2}
	{\sum_{j\neq k}\big|\mathbf{h}_k^{\s{H}}(t)\mathbf{v}_j(t)\big|^2 + \sigma_0^2}
	\right).
\end{equation}

\subsubsection{Multimodal Sensing Model}
To facilitate robust communication and precise environmental perception within dynamic low-altitude scenarios, the multimodal ISAC framework synergizes heterogeneous sensing modalities, capturing complementary environmental features that augment the limited spatial awareness provided by standard RF communication signals.
Formally, at time slot $t$, the system aggregates a set of heterogeneous sensing inputs, denoted as
\begin{equation}
	\mathbf{S}(t) = \{\mathbf{S}_{\textup{vis}}(t), \mathbf{S}_{\textup{radar}}(t), \mathbf{S}_{\textup{lidar}}(t), \mathbf{S}_{\textup{pos}}(t)\},
\end{equation}
where $\mathbf{S}_{\textup{vis}}(t), \mathbf{S}_{\textup{radar}}(t), \mathbf{S}_{\textup{lidar}}(t)$, and $ \mathbf{S}_{\textup{pos}}(t)$ denote the sensing data from the vision, radar, lidar, and position modalities, respectively.
These data are obtained from sensors either mounted on the UAVs or co-located at the ground BS.
While $\mathbf{S}(t)$ is defined as a superset of potential sensing sources, our framework is inherently flexible and can operate with available subset of these modalities, depending on specific hardware configurations or instantaneous sensor availability.
The specific characteristics and data representations of each modality are detailed as follows.

\begin{itemize}[leftmargin=*]
	\item Visual modality $\mathbf{S}_{\textup{vis}}(t)$: The vision modality leverages RGB images to capture rich semantic and contextual information, such as the spatial distribution of obstacles and dynamic targets. Such visual data provide critical environmental awareness for downstream ISAC tasks like beam prediction and trajectory tracking.
	To optimize learning efficiency, the raw visual data is preprocessed through spatial filtering and background subtraction to suppress static environmental noise and isolate dynamic regions of interest (e.g., moving UAVs), thereby distilling the most task-relevant features for subsequent inference.
	\item Radar modality $\mathbf{S}_{\textup{radar}}(t)$: The radar modality leverages Doppler signatures to provide precise measurements of target distance and relative velocity.
	Its primary advantage lies in the inherent robustness against adverse environmental conditions, such as fog, rain, or low light conditions, where optical sensors may fail.
	By encapsulating these kinematic characteristics, this modality offers critical motion information that enables the ISAC system to accurately track dynamic targets and anticipate mobility patterns even in visually degraded scenarios.
	\item Lidar modality $\mathbf{S}_{\textup{lidar}}(t)$: Providing high-resolution three-dimensional geometric information, the lidar modality captures precise depth perception and spatial structural details that two-dimensional visual sensors inherently lack.
	By offering a fine-grained volumetric representation of the environment, this modality facilitates accurate spatial reasoning, enabling the system to resolve complex object shapes and determine precise physical locations essential for environment-aware communication and sensing tasks.
	\item Position modality $\mathbf{S}_{\textup{pos}}(t)$: Providing a macroscopic spatial reference, the position modality supplies absolute localization coordinates and long-term motion trends. Unlike local environmental sensors, the position data offers a global context for user movement, serving as a critical coarse-grained information to anchor spatial relationships and guide decision-making in large-scale dynamic scenarios.
	
\end{itemize}
Collectively, the integration of these sensing modalities enables the construction of a comprehensive and task-relevant environmental representation. This multi-dimensional awareness is pivotal for supporting the dual objectives of robust communication and precise sensing in dynamic low-altitude environments.

\subsection{Problem Formulation}

To achieve robust perception and connectivity in dynamic LAWNs, we formulate a unified multimodal learning framework that models the interaction between heterogeneous sensing modalities and RF communication signals as a generalized cross-modal prediction problem.
This formulation unifies diverse ISAC tasks under a common optimization paradigm, enabling mutual performance enhancement by utilizing environmental perception to optimize connectivity or leveraging wireless signals to improve sensing precision.

Formally, at each time slot $t$, the multimodal ISAC system aggregates heterogeneous sensing inputs $\mathbf{S}(t)$ from different sensing modalities such as vision, radar, and lidar. In parallel, it incorporates RF link measurements $\mathbf{C}(t)$, which encompass wireless metrics including historical beam indices, received signal strength indicators, and CSI.
To accommodate diverse task requirements, we define a unified system input $\mathbf{X}(t)$ by selectively aggregating heterogeneous sensing inputs $\mathbf{S}(t)$ and RF link measurements $\mathbf{C}(t)$.
The resulting input space is defined as
\begin{equation}
	\mathbf{X}(t) \in \left\{ \mathbf{S}(t),\, \mathbf{C}(t),\, \left\{ \mathbf{S}(t), \mathbf{C}(t) \right\} \right\},
\end{equation}
which allows the predictive model to operate on either individual domains or a composite measurement.
We aim to learn a parametric mapping function $g_{\boldsymbol{\Theta}}$ that maps the system input $\mathbf{X}(t)$ to a task-specific output $\mathbf{o}(t)$, such as a discrete beam index or a continuous UAV coordinate. The learning objective is formulated as an expected risk minimization problem, where we seek the optimal parameters $\boldsymbol{\Theta}$ that minimize the expected loss $\mathcal{L}\left( \mathbf{\hat o}(t),\, \mathbf{o}(t) \right)$ between the model output $\mathbf{\hat o}(t)$ and the corresponding ground truth $ \mathbf{o}(t)$ as follows,
\begin{equation}
	\begin{aligned}\label{general_formulation}
		\min_{\boldsymbol{\Theta}} &~~ \mathbb{E} \left[ \mathcal{L}\left( \mathbf{\hat o}(t),\, \mathbf{o}(t) \right) \right], \\
		\textup{s.t.} &~~ \mathbf{\hat o}(t) = g_{\boldsymbol{\Theta}}(\mathbf{X}(t)),
	\end{aligned}
\end{equation}
where the expectation is taken over the spatiotemporal variability of the environment, channel states, and sensor observations.

This generalized formulation provides a versatile foundation applicable to a broad spectrum of ISAC scenarios. By customizing the output space and loss function, the framework seamlessly applies to diverse problem paradigms. For instance, defining $\mathbf{o}(t)$ as a beam index under cross-entropy loss recovers the beam prediction formulation, while setting $\mathbf{o}(t)$ as a position vector under mean squared error (MSE) yields the UAV trajectory tracking problem.
To demonstrate this adaptability, we formally instantiate this framework through three representative ISAC tasks.

\subsubsection{Sensing-Aided Channel Estimation}

Precise acquisition of CSI is pivotal for mmWave communications but poses significant challenges in LAWNs due to the channel variations caused by rapid mobility.
To mitigate the overhead of dense pilot signaling, we leverage heterogeneous multimodal sensing data $\mathbf{S}(t)$ as auxiliary side information to estimate the downlink channel vector $\mathbf{h}_k(t)$.
To specialize the general Problem \eqref{general_formulation} for this task, we instantiate the system input $\mathbf{X}(t)$ as the multimodal sensing data $\mathbf{S}(t)$ and the model output $\mathbf{\hat o}(t)$ as the estimated downlink channel vector $\mathbf{\hat h}_k(t)$, i.e.,
\begin{equation}
	\mathbf{X}(t) \triangleq \mathbf{S}(t) \quad \text{and} \quad \mathbf{\hat o}(t) \triangleq \mathbf{\hat h}_k(t).
\end{equation}
Consequently, the optimization objective is concretized as a regression problem that minimizes the expected MSE between the estimated channel \(\mathbf{\hat{h}}_k(t)\) and the ground truth \(\mathbf{h}_k(t)\), given by
\begin{equation}
	\begin{aligned}
		\min_{\boldsymbol{\Theta}} &~~ \mathbb{E}\left[  \left\| \mathbf{h}_k(t) - \mathbf{\hat h}_k(t) \right\|^2  \right] \\
		\textup{s.t.}\,& ~~\mathbf{\hat h}_k(t) = g_{\boldsymbol{\Theta}}(\mathbf{S}(t)).
	\end{aligned}
\end{equation}
By exploiting the rich spatial and semantic context provided by UAV-mounted and BS-mounted sensors, this formulation effectively captures the implicit correlations between environmental features and signal propagation, thereby enhancing CSI estimation accuracy while reducing the dependence on conventional pilot-based methods.

\subsubsection{Sensing-Aided Beamforming Design}

Efficient beam management is paramount for UAV-enabled mmWave communication systems, yet traditional beam sweeping incurs prohibitive latency overheads given the highly directional transmission and rapid aerial mobility.
To enable low-latency beam management, we instantiate the unified Problem \eqref{general_formulation} as a proactive beam prediction task, which leverages heterogeneous sensing data to directly predict the optimal beamforming configuration.
Specifically, the system input $\mathbf{X}(t)$ and output $\mathbf{\hat o}(t)$ are respectively instantiated as the multimodal sensing data $\mathbf{S}(t)$ and the estimated beamforming vectors $\{\mathbf{\hat{v}}_k(t)\}_{k=1}^K$ for all UAVs, i.e.,
\begin{equation}
	\mathbf{X}(t) \triangleq \mathbf{S}(t) \quad \text{and} \quad \mathbf{\hat o}(t) \triangleq \{\mathbf{\hat{v}}_k(t)\}_{k=1}^K.
\end{equation}
The general learning objective is specialized to maximize the instantaneous downlink sum-rate as follows,
\begin{equation}
	\begin{aligned}
		\max_{\boldsymbol{\Theta}} &~~ \mathbb{E} \left[ \sum_{k=1}^{K} 
		\log_2 \left( 1 + \frac{|\mathbf{h}_k^H(t)\mathbf{\hat{v}}_k(t)|^2}
		{\sum_{j\neq k} |\mathbf{h}_k^H(t)\mathbf{\hat{v}}_j(t)|^2 + \sigma_0^2} \right) \right] \\
		\textup{s.t.} &~~ \mathbf{\hat{v}}_k(t) = g_{\boldsymbol{\Theta}}(\mathbf{S}(t)), \forall k,
	\end{aligned}
\end{equation}
where \(\mathbf{h}_k(t)\) denotes the downlink channel vector between the BS and UAV \(k\), and \(\sigma_0^2\) is the noise power.
By bypassing explicit channel estimation and exhaustive search, this approach effectively leverages the spatial intelligence embedded in multimodal observations to achieve accurate, mobility-aware beam selection, thereby stabilizing connectivity against the dynamic fluctuations of the low-altitude channel.

\subsubsection{Communication-Aided UAV Trajectory Tracking}

Accurate trajectory tracking is indispensable for maintaining reliable connectivity in LAWNs, yet relying solely on onboard sensors often leads to the degradation of localization accuracy due to signal blockage or inertial drift.
To address this challenge, we employ onboard sensing data $\mathbf{S}(t)$ with downlink RF measurements $\mathbf{C}(t)$ to achieve high-precision positioning.
Specifically, the system input $\mathbf{X}(t)$ and output $\mathbf{\hat o}(t)$ in Problem \eqref{general_formulation} are respectively instantiated as the joint sensing and RF measurements $\{ \mathbf{S}(t), \mathbf{C}(t) \}$ and estimated UAV position ${ \mathbf{\hat p}(t) }$, i.e.,
\begin{equation}
	\mathbf{X}(t) \triangleq \{ \mathbf{S}(t), \mathbf{C}(t) \} \quad \text{and} \quad \mathbf{\hat o}(t) \triangleq { \mathbf{\hat p}(t) }.
\end{equation}
The optimization objective is concretized as a trajectory regression problem that minimizes the cumulative position error over the time horizon as follows,
\begin{equation}
	\begin{aligned}
		\min_{\boldsymbol{\Theta}} &~~ \mathbb{E} \left[ \sum_{\tau=t-L}^{t} \left\| \mathbf{p}(\tau) - \hat{\mathbf{p}}(\tau) \right\|^2 \right] \\
		\textup{s.t.} &~~ \hat{\mathbf{p}}(\tau) = g_{\boldsymbol{\Theta}}(\{   \mathbf{S}(\tau), \mathbf{C}(\tau)\}), \forall \tau \in [t-L, t],
	\end{aligned}
\end{equation}
where $\mathbf{p}(\tau)$ represents the ground-truth coordinates at time slot $\tau$.
By integrating these sources, the system utilizes RF observations to correct long-term sensor drift while leveraging sensing data to sustain trajectory estimation during signal interruptions, ensuring reliable localization in low-altitude environments.

%

\subsection{Existing Approaches}

Multimodal data fusion in ISAC systems has evolved through various paradigms. In this subsection, we categorize representative methods into two primary classes, namely static fusion strategies \cite{10912462} and monolithic neural architectures \cite{10539181, 11208675, 10829757}, and highlight their respective limitations to motivate the design of our proposed method.

\subsubsection{Static fusion strategies}
The most straightforward approach to multimodal fusion relies on fixed rules or hand-crafted heuristics. Formally, let $D$ denote the total number of modalities. For each modality $d$, an intermediate feature embedding \(\mathbf{z}_d(t)\) is first extracted from the raw input. These modality-specific embeddings are then aggregated into a joint representation \(\mathbf{z}(t)\) via deterministic operations:
\begin{itemize}
	\item {Direct Concatenation \cite{10912462}:}
	Feature vectors from different modalities are concatenated into a single high-dimensional vector
	\begin{equation}
		\mathbf{z}(t) = [\mathbf{z}_1(t)^{\s{T}}, \ldots, \mathbf{z}_D(t)^{\s{T}}]^{\s{T}},
	\end{equation}
	which is subsequently fed into a downstream predictor.
	
	\item {Weighted Summation:}
	Features are combined using pre-assigned scalar weights as follows,
	\begin{equation}
		\mathbf{z}(t) = \sum_{d=1}^{D} w_d \mathbf{z}_d(t),
	\end{equation}
	where \(w_d\) is typically determined by expert knowledge or heuristic confidence estimates.
\end{itemize}
Although computationally efficient and easy to implement, such static strategies fuse all inputs using pre-assigned and fixed weights, regardless of their instantaneous quality.
This limitation is particularly critical in dynamic LAWNs, where sensor reliability fluctuates rapidly due to blockage or adverse environmental conditions. Consequently, static fusion strategies fail to filter out degraded sensing data, inevitably leading to performance degradation.

\subsubsection{Monolithic neural architectures}

To overcome the limitations of static fusion, recent research has shifted toward unified deep learning frameworks, such as multimodal transformers \cite{10539181, 11208675} and MLLMs \cite{10829757}.
These approaches employ a monolithic architecture where heterogeneous inputs are projected onto a shared latent space and processed through a dense backbone to capture cross-modal correlations. However, this design presents critical bottlenecks for the practical ISAC deployment in LAWNs.
First, sharing parameters across modalities often leads to modality interference, where dominant sensors (e.g., vision) suppress the learning of others, reducing robustness when the primary source fails. Second, these models enforce unconditional computation, activating the full network for every inference step regardless of data quality. This processing overhead creates a significant energy burden for battery-powered UAVs, hindering deployment in resource-constrained aerial scenarios.

\begin{figure*}[t]
	\renewcommand\figurename{Fig.}
	\centering \vspace*{0pt} \setlength{\baselineskip}{10pt}
	\includegraphics[width = 1\textwidth,trim=0 0 0 0,clip]{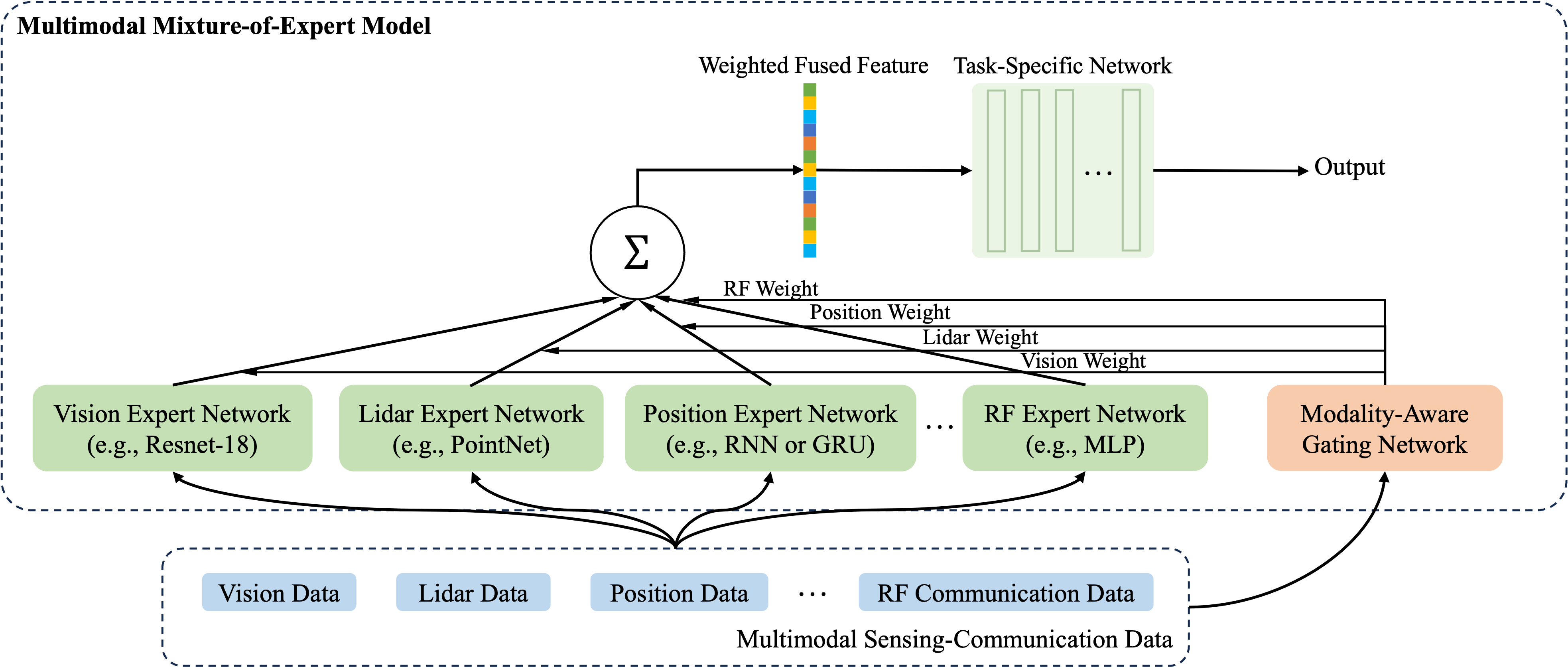}
	\caption{Architecture of the proposed multimodal MoE framework. Modality-specific experts extract features from heterogeneous inputs, which are aggregated via weighted summation using coefficients from a gating network to form a fused representation for the prediction head.}\label{fig_moe}
\end{figure*}

In summary, static fusion strategies are computationally efficient but rely on fixed fusion rules that cannot adapt to changing modality quality or dynamic environments.
In contrast, monolithic neural architectures that process all modalities jointly offer strong representation power but incur high computational cost and cross-modal interference because all modality-specific branches are always activated, even when some modalities are noisy or uninformative.
To address both the lack of adaptivity in static fusion and the high cost of monolithic neural architectures, we propose a multimodal MoE framework in the following section.
Instead of a single monolithic network, our approach uses a modular architecture in which specialized expert networks are coordinated by a gating mechanism that adaptively weights experts based on the current input.
This design effectively combines environmental adaptability with computational efficiency, enabling the system to intelligently prioritize reliable sensing modalities for robust perception while minimizing the energy burden on resource-constrained aerial platforms.

\section{Proposed Multimodal MoE Framework}

In this section, we present the proposed multimodal MoE framework, specifically designed to address the dynamic requirements of ISAC in LAWNs.
As illustrated in Fig.~\ref{fig_moe}, the architecture consists of modality-specific expert networks and a modality-aware gating network.
We first detail the mathematical formulation of these components, followed by the description of the end-to-end training procedure.

\subsection{Multimodal MoE Architecture}

To effectively extract discriminative features from heterogeneous modalities, the proposed framework constructs a set of parallel expert networks. Unlike monolithic architectures that enforce a unified processing path, our design adopts a modular approach where each sensing modality is handled by a dedicated expert network.
We define the $d$-th expert as a non-linear mapping function $g_d(\cdot;\boldsymbol{\theta}_d)$, parameterized by the trainable weights $\boldsymbol{\theta}_d$. The complete ensemble of experts is denoted as $\mathcal{E}=\{g_d(\cdot;\boldsymbol{\theta}_d)\}_{d=1}^{D}$. This decoupled structure allows each expert to employ a distinct inductive bias tailored to the specific data format of its corresponding modality.
For a given time slot $t$, let $\mathbf{X}(t)=\{\mathbf{X}_d(t)\}_{d=1}^{D}$ represent the set of synchronized multimodal inputs. Each expert processes its specific input $\mathbf{X}_d(t)$ to generate the high-dimensional feature embedding $\mathbf{z}_{d}(t)$ as
\begin{equation}
	\mathbf{z}_{d}(t)=g_d(\mathbf{X}_d(t);\boldsymbol{\theta}_d), \forall d.
\end{equation}

To improve the feature representation for downstream ISAC tasks, the neural architecture of each expert is specifically customized to align with the characteristics of its corresponding modality.
For instance, since visual images are rich in spatial patterns and object details, the visual expert typically utilizes powerful feature extractors such as ResNet \cite{he2016deep} or vision transformers \cite{dosovitskiy2020vit} to interpret complex semantic context. In contrast, radar signals often contain time-varying motion information. To capture these dynamic attributes including target range and velocity, the radar expert employs architectures suited for sequential or spectral processing, such as convolutional neural networks \cite{krizhevsky2012imagenet} or long short-term memory models \cite{hochreiter1997long}. Furthermore, handling lidar data requires a different approach because they consist of scattered and unordered three-dimensional points rather than a regular image grid. Consequently, the lidar expert leverages specialized geometric networks including PointNet \cite{qi2017pointnet} or voxel-based methods \cite{deng2021voxel} to accurately reconstruct spatial depth and structure. This tailored design ensures that the unique strengths of each sensing modality are fully preserved and utilized in the final fusion process.

%
%
%

While independent feature extraction effectively captures modality-specific patterns, relying solely on isolated processing neglects the complex interactions between heterogeneous data modalities. In practical deployment over LAWNs, the reliability of each sensing modality fluctuates dynamically due to changing environmental conditions and sensor limitations. Consequently, fusion strategies that assign static or uniform weights often lead to suboptimal performance when some modalities are corrupted. To overcome this limitation, it is critical to equip the framework with an adaptive mechanism capable of assessing the instantaneous contribution of each expert. Motivated by this, we propose a modality-aware gating network that dynamically evaluates the relative importance of diverse inputs to generate context-aware fusion weights, as detailed in the subsequent subsection.

\subsection{Modality-Aware Gating Network}
The modality-aware gating network functions as the central decision module within the proposed multimodal MoE framework.
While the expert networks focus on extracting deep semantic features from each modality, the gating network is tasked with evaluating the global context to determine the instantaneous reliability of each expert. In dynamic low-altitude environments, factors such as variable lighting, blockage, or signal clutter can cause the informativeness of individual modalities to fluctuate rapidly.
Consequently, static fusion strategies that assign fixed weights fail to adapt to these variations and exhibit performance degradation when a primary sensor is compromised.
To overcome this limitation, the gating network learns to model the uncertainty associated with each modality and dynamically generates a probability distribution over the experts, effectively prioritizing reliable signals while suppressing noisy or irrelevant inputs.

Formally, the gating network is defined as a learnable non-linear mapping $g_{\textup{gate}}(\cdot;\boldsymbol{\theta}_{\textup{gate}})$ parameterized by $\boldsymbol{\theta}_{\textup{gate}}$.
It accepts the synchronized multimodal input set $\mathbf{X}(t)$ as context and computes a vector of normalized gating coefficients. These coefficients represent the relative contribution of each modality at time slot $t$ and are calculated via the softmax function as
\begin{equation}
	[w_{1}(t), w_{2}(t), \dots, w_{D}(t)] = \text{softmax}(g_{\textup{gate}}(\mathbf{X}(t);\boldsymbol{\theta}_{\textup{gate}})),
\end{equation}
where the softmax activation ensures that the generated weights are non-negative and sum to one, i.e.,
\begin{equation}
	\begin{aligned}
		&\sum_{d=1}^{D} w_d(t) = 1, \\
		 &w_d(t)\geq 0, \forall d.
	\end{aligned}
\end{equation}

In terms of architecture, the gating network is designed to be computationally lightweight to minimize inference latency. It typically comprises a multilayer perceptron (MLP) with non-linear activation functions. The processing flow begins by aggregating the heterogeneous inputs into a shared latent embedding to capture cross-modal dependencies and environmental context. This compact representation is subsequently projected into a vector of raw logits, which are then normalized to produce the final fusion weights. Furthermore, the specific architecture of the gating network is adaptable to the complexity of the input data modality. For instance, when processing high-dimensional inputs like visual images or radar maps, the network may incorporate lightweight convolutional layers or global average pooling operations to extract global context descriptors before the dense gating layers. This flexible design enables accurate assessment of modality relevance without imposing a significant computational burden.

Upon obtaining the normalized gating weights, the final fused multimodal representation $\mathbf{z}(t)$ is computed as a weighted aggregation of the modality-specific expert embeddings $\{\mathbf{z}_d(t)\}_{d=1}^D$ as follows,
\begin{equation}
	\mathbf{z}(t) = \sum_{d=1}^{D} w_d(t)\,\mathbf{z}_{d}(t).
\end{equation}
This aggregated feature vector captures complementary information extracted from heterogeneous sensing modalities and serves as the input to the task-specific prediction head. Formally, we define the prediction head as a non-linear mapping $g_o(\cdot;\boldsymbol{\theta}_o)$, parameterized by $\boldsymbol{\theta}_o$, which transforms the fused representation $\mathbf{z}(t)$ into the final task-specific prediction $\hat{\mathbf{o}}(t)$ as follows,
\begin{equation}
	\hat{\mathbf{o}}(t) = g_{o}(\mathbf{z}(t);\boldsymbol{\theta}_{o}).
\end{equation}
The architecture of the prediction head $g_o(\cdot;\boldsymbol{\theta}_o)$ is customized to the specific nature of the downstream ISAC task:
\begin{itemize} \item For classification tasks, such as predicting the optimal beam index, the network ends with a softmax layer that outputs a probability distribution across all possible categories and is trained using a cross-entropy loss function to select the most likely candidates. \item For regression tasks, such as estimating channel parameters or tracking UAV trajectories, the network ends with linear output units. This configuration directly predicts continuous numerical values and is optimized by minimizing the error between the predicted and actual values, typically using MSE. \end{itemize}

In summary, by jointly optimizing the modality-specific experts $\{g_d(\cdot;\boldsymbol{\theta}_d)\}_{d=1}^D$, the adaptive gating network $g_{\textup{gate}}(\cdot;\boldsymbol{\theta}_{\textup{gate}})$, and the prediction head $g_o(\cdot;\boldsymbol{\theta}_o)$, the proposed multimodal MoE framework establishes a robust end-to-end learning paradigm for multimodal ISAC.
This mechanism enables the system to dynamically balance modality contributions based on real-time environmental contexts, thereby significantly enhancing predictive accuracy and resilience against sensor uncertainties in dynamic low-altitude scenarios.

\subsection{Algorithm Development}

In this subsection, we present the training procedure for the proposed multimodal MoE framework in detail. Specifically, our objective is to jointly optimize the parameters of the expert networks, the modality-aware gating network, and the subsequent prediction head. To this end, we formulate the training as a general supervised learning problem using a labeled multimodal dataset.

Given a training dataset of synchronized multimodal inputs and ground-truth labels, denoted as $\mathcal{D}=\{(\mathbf{X}(t), \mathbf{o}(t))\}_{t=1}^{T}$, our goal is to minimize the empirical risk as measured by the loss function $\mathcal{L}(\cdot)$.
Depending on the specific ISAC task, the loss function $\mathcal{L}(\cdot)$ can be chosen as binary cross-entropy (BCE) for classification tasks (e.g., proactive blockage prediction), MSE for regression tasks (e.g., channel estimation), or negative sum-rate loss for beamforming optimization tasks.
Formally, the complete parameter optimization problem can be expressed as
\begin{equation}\label{dense_problem}
	\min_{\{\boldsymbol{\theta}_d\}_{d=1}^D,\boldsymbol{\theta}_{\textup{gate}},\boldsymbol{\theta}_o}~\frac{1}{T}\sum_{t=1}^{T}\mathcal{L}\bigl(\mathbf{\hat o}(t), {\mathbf{o}}(t)\bigr),
\end{equation}
where the prediction output $\hat{\mathbf{o}}(t)$ is obtained by sequentially computing expert outputs, modality-aware fusion weights, fused representations, and final predictions as follows,
\begin{equation}
\begin{aligned}
	\mathbf{z}_{d}(t)&=g_d(\mathbf{X}_d(t);\boldsymbol{\theta}_d), ~\forall d,\\[3pt]
	[w_{1}(t),\dots,w_{D}(t)]&=\text{softmax}(g_{\textup{gate}}(\mathbf{X}(t);\boldsymbol{\theta}_{\textup{gate}})),\\[3pt]
	\mathbf{z}(t)&=\sum_{d=1}^{D}w_d(t)\mathbf{z}_d(t),\\[3pt]
	\hat{\mathbf{o}}(t)&=g_o(\mathbf{z}(t);\boldsymbol{\theta}_o).
\end{aligned}
\end{equation}

\begin{algorithm}[!tbp]
	\caption{Training Procedure of the Proposed Multimodal MoE Framework.}
	\label{algo:multimodal_moe}
	\KwIn{Training dataset $\mathcal{D}=\{(\mathbf{X}(t),\mathbf{o}(t))\}_{t=1}^{T}$, initialized expert network parameters $\{\boldsymbol{\theta}_d\}_{d=1}^{D}$, gating network parameters $\boldsymbol{\theta}_{\textup{gate}}$, prediction head parameters $\boldsymbol{\theta}_o$, learning rate $\eta$.}
	
	\For{epoch $=1,2,\dots, E$}{
		Shuffle the training dataset $\mathcal{D}$\;
		\For{each training sample $(\mathbf{X}(t),\mathbf{o}(t))\in\mathcal{D}$}{
			Compute modality-specific expert embeddings:
			\vspace{-2pt}
			\begin{equation}
				\vspace{-2pt}
				\begin{aligned}
					\mathbf{z}_{d}(t)=g_d(\mathbf{X}_d(t);\boldsymbol{\theta}_d),~\forall d;
				\end{aligned}
			\end{equation}
			
			Compute modality-aware fusion weights:
			\vspace{-2pt}
			\begin{equation}
				\begin{aligned}
					[w_{1}(t),\dots,w_{D}(t)]=\text{softmax}(g_\textup{gate}(\mathbf{X}(t);\boldsymbol{\theta}_\textup{gate}));
				\end{aligned}
			\end{equation}
			
			Compute fused multimodal representation:
			\vspace{-4pt}
			\begin{equation}
				\begin{aligned}
					\mathbf{z}(t)=\sum_{d=1}^{D} w_d(t)\mathbf{z}_{d}(t);
				\end{aligned}
			\end{equation}
			
			Compute final prediction output:
			\vspace{-2pt}
			\begin{equation}
				\begin{aligned}
					\hat{\mathbf{o}}(t)=g_o(\mathbf{z}(t);\boldsymbol{\theta}_o);
				\end{aligned}
			\end{equation}
			
			Evaluate supervised loss function:
			\vspace{-2pt}
			\begin{equation}
				\vspace{-1pt}
				\begin{aligned}
					\mathcal{L}_t=\mathcal{L}(\mathbf{o}(t),\hat{\mathbf{o}}(t));
				\end{aligned}
			\end{equation}
			
			Compute gradients via backpropagation:
			\vspace{-2pt}
			\begin{equation}
				\vspace{-1pt}
				\begin{aligned}
					\nabla_{\boldsymbol{\theta}_d}\mathcal{L}_t,\forall d;\quad\nabla_{\boldsymbol{\theta}_\textup{gate}}\mathcal{L}_t;\quad\nabla_{\boldsymbol{\theta}_o}\mathcal{L}_t;
				\end{aligned}
			\end{equation}
			
			Update parameters with gradient descent:
			\vspace{-2pt}
			\begin{equation}
				\vspace{-3pt}
				\begin{aligned}
					\boldsymbol{\theta}_d&\leftarrow\boldsymbol{\theta}_d-\eta\nabla_{\boldsymbol{\theta}_d}\mathcal{L}_t,~\forall d ;\\
					\boldsymbol{\theta}_\textup{gate}&\leftarrow\boldsymbol{\theta}_\textup{gate}-\eta\nabla_{\boldsymbol{\theta}_\textup{gate}}\mathcal{L}_t;\\
					\boldsymbol{\theta}_o&\leftarrow\boldsymbol{\theta}_o-\eta\nabla_{\boldsymbol{\theta}_o}\mathcal{L}_t;
				\end{aligned}
			\end{equation}
		}
	}
	\KwOut{Expert parameters $\{\boldsymbol{\theta}_d\}_{d=1}^{D}$, gating parameter $\boldsymbol{\theta}_\textup{gate}$, and prediction head parameter $\boldsymbol{\theta}_o$.}
\end{algorithm}

The complete end-to-end training procedure is formally summarized in Algorithm \ref{algo:multimodal_moe}. By minimizing the task-specific loss $\mathcal{L}(\cdot)$ via backpropagation, the framework jointly optimizes the expert backbones, the gating mechanism, and the prediction head within a unified computational graph. This joint training process allows the model to learn a context-aware fusion policy that dynamically recalibrates modality weights based on the instantaneous environmental state.
As a result, the trained multimodal MoE framework tends to exploit information from reliable sensing modalities while reducing the influence of degraded ones, leading to robust performance across diverse ISAC applications in dynamic low-altitude environments.

\section{Sparse Multimodal MoE Framework}
\label{sec:sparse_moe}

In this section, we propose a sparse multimodal MoE framework to meet the stringent energy and computation constraints of the aerial platforms in LAWNs. By selectively activating only the top-$N$ most relevant experts via a sparse routing mechanism, this variant avoids unnecessary processing and reduces resource overhead while maintaining adaptive fusion.

\begin{figure}[t]
	\centering
	\includegraphics[width=0.5\textwidth,trim=10 20 10 0,clip]{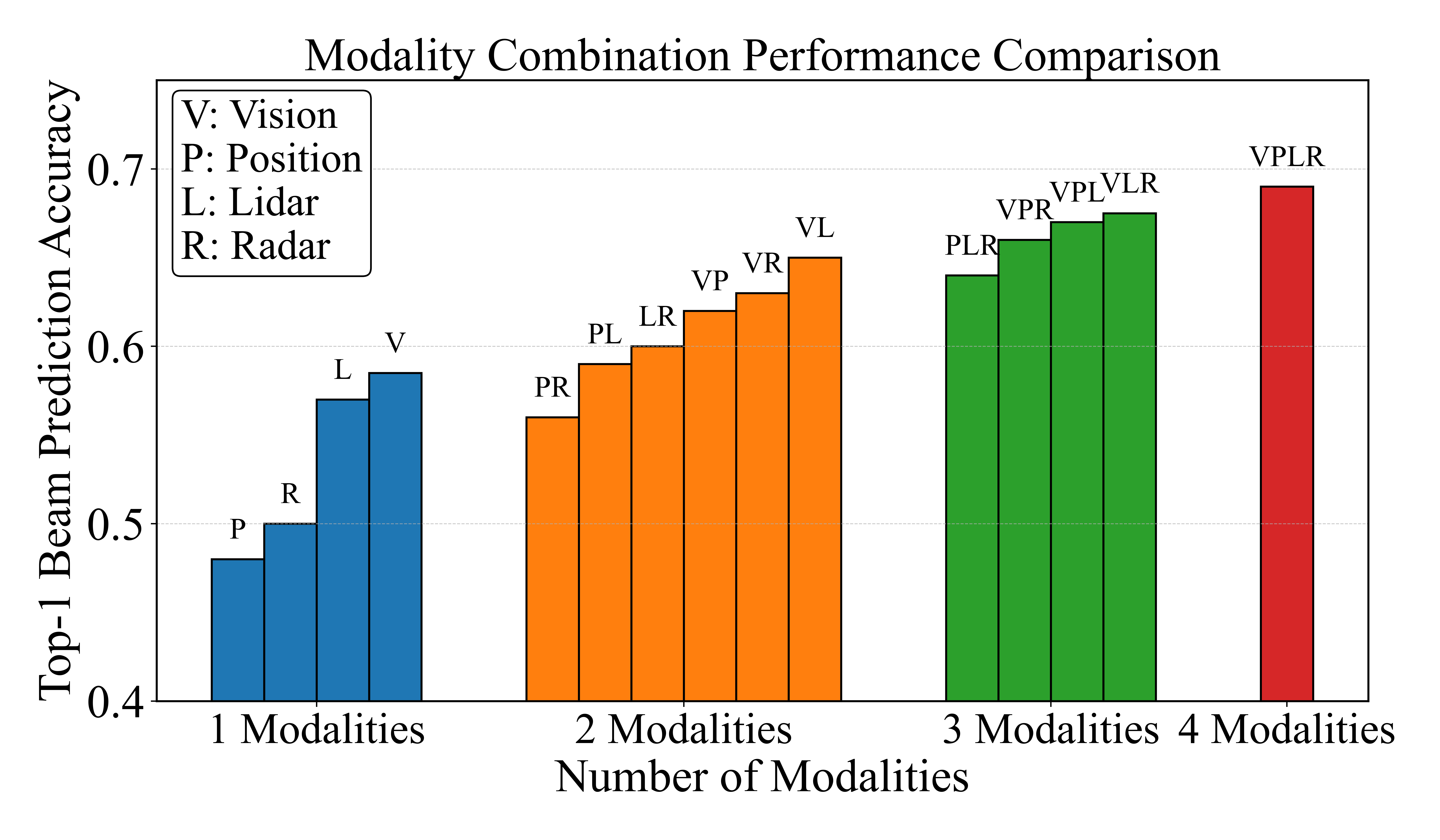}
	\caption{Top-1 beam prediction accuracy across different combinations of sensing modalities.}
	\label{fig:sparse_motivation}
\end{figure}

\subsection{Sparse Expert Activation}

While the proposed multimodal MoE framework offers comprehensive fusion, activating all $D$ experts for every inference step imposes a computational burden that challenges the energy constraints of aerial platforms. However, as evidenced by the performance comparison in Fig.~\ref{fig:sparse_motivation}, the accuracy gain from integrating additional modalities exhibits clear diminishing returns. Specifically, the marginal improvement decreases significantly as the number of modalities grows (e.g., the gap between 3 and 4 modalities is minimal), implying that fully activating the entire expert ensemble is often unnecessary and computationally redundant.

To capitalize on this observation, we propose a sparse multimodal MoE framework that adaptively activates different experts.
By dynamically selecting only the top-$N$ most informative experts per input (with $N < D$), this approach effectively prunes redundant branches.
This yields a favorable trade-off between inference accuracy and computational efficiency for real-time aerial applications.
Specifically, the sparse MoE shares the same expert pool and gating network as the dense model. At each time slot \(t\), the gating network first produces a set of fusion scores \([w_1(t), \dots, w_D(t)]\) as in the dense case. Instead of using all experts, the sparse variant only activates those associated with the largest scores. Concretely, we select the indices of the top-$N$ experts that are most relevant for the current multimodal observation as
\begin{equation}
	\mathcal{S}(t) = \text{TopN}\big(w_1(t), \dots, w_D(t)\big),
\end{equation}
where $\text{TopN}(\cdot)$ returns the indices of the $N$ largest entries among $\{w_d(t)\}_{d=1}^{D}$. Based on this set, we define a binary activation mask to indicate whether the $d$-th expert is activated at time slot $t$, given by
\[
m_d(t) = 
\begin{cases}
	1, & \text{if } d \in \mathcal{S}(t), \\
	0, & \text{otherwise},
\end{cases}
\quad d=1,\dots,D.
\]
In the multimodal MoE proposed in Section III, all expert embeddings $\{\mathbf{z}_d(t)\}_{d=1}^{D}$ are computed and combined according to the normalized weights. In contrast, the sparse MoE only evaluates the experts with $m_d(t)=1$. That is, for each time slot $t$, we compute
\[
\mathbf{z}_d(t) = g_d(\mathbf{X}_d(t);\boldsymbol{\theta}_d)\quad \text{only if } m_d(t)=1,
\]
and skip the forward pass through the remaining experts, which directly leads to a reduction in computational cost and memory access. To ensure that the fused representation remains a convex combination of the active expert outputs, we renormalize the original weights over the selected subset as
\begin{equation}
	\tilde{w}_d(t) = \frac{m_d(t) w_d(t)}{\sum_{j=1}^{D} m_j(t) w_j(t) + \varepsilon}, \quad d=1,\dots,D,
\end{equation}
where \(\varepsilon > 0\) is a small constant added for numerical stability. By construction, $\tilde{w}_d(t)\ge 0$ and $\sum_{d=1}^{D}\tilde{w}_d(t)=1$, and $\tilde{w}_d(t)=0$ for all inactive experts with $m_d(t)=0$.
With these sparsified weights, the fused multimodal representation is given by
\begin{equation}
	\mathbf{\tilde z}(t) = \sum_{d=1}^{D} \tilde{w}_d(t)\,\mathbf{z}_d(t),
\end{equation}
where the sum effectively runs only over the $N$ active experts.
Compared to the dense case, the sparse MoE preserves the gating behavior, since larger $w_d(t)$ still leads to larger normalized weights $\tilde w_d(t)$, while reducing computation and memory cost from evaluating all $D$ experts to only $N < D$ experts per time slot.

\subsection{Training Procedure with Sparse Activation}

The integration of the top-$N$ selection operator introduces a non-differentiable step into the computational graph, which blocks the flow of gradients required for standard backpropagation. To circumvent this issue and enable end-to-end optimization, we employ the straight-through estimator (STE) \cite{bengio2013estimating}. This technique decouples the forward execution from the backward gradient approximation, allowing the network to make hard routing decisions while remaining differentiable for parameter updates.

During the forward propagation phase, the system performs selective execution, activating only the experts deemed most relevant.
The gating network generates the raw probability vector $\mathbf{w}(t)$, which is then processed by the non-differentiable top-$N$ operator to yield a binary mask and the renormalized sparse weights $\mathbf{\tilde{w}}(t)$. Consequently, only the selected subset of experts contributes to the fused representation $\mathbf{z}(t)$, improving computational efficiency.
In contrast, during the backward propagation phase, the STE heuristic approximates the hard top-$N$ operation as an identity function for the purpose of gradient calculation. Specifically, we assume that the gradients flow through the discrete selection layer without attenuation. Formally, the gradient of the loss function $\mathcal{L}$ with respect to the original gating weights $\mathbf{w}(t)$ is approximated by the gradients computed for the sparsified weights $\mathbf{\tilde{w}}(t)$ as follows,
\begin{equation}
	\frac{\partial \mathcal{L}}{\partial \mathbf{w}(t)} \approx \frac{\partial \mathcal{L}}{\partial \mathbf{\tilde{w}}(t)}.
\end{equation}
This approximation implies that if an expert is activated, its gradient contribution is backpropagated directly to the gating network. If an expert is inactive (i.e., $\tilde{w}_d(t)=0$), it receives no gradient update for that instance. This mechanism allows the gating parameter $\boldsymbol{\theta}_g$ to be updated via stochastic gradient descent despite the discrete nature of the inference process.
Under this training paradigm, the overall learning objective remains consistent with the empirical risk minization problem in \eqref{dense_problem}.
By leveraging the STE, the sparse MoE framework effectively learns an adaptive and sparsity-constrained routing policy that balances prediction accuracy with the stringent resource limitations in practical LAWNs.

\subsection{Computational Benefits and Practical Considerations}

The sparse MoE framework inherits the architectural flexibility of the multimodal MoE in Section III, while achieving significantly improved computational efficiency. Given that only $N$ experts are evaluated per input (with $N < D$), the total computational cost per forward pass is reduced from $\mathcal{O}(D)$ to $\mathcal{O}(N)$. The same reduction applies to memory and energy usage, making the sparse MoE model particularly attractive for deployment on embedded processors and real-time UAV platforms.
Additionally, by enforcing sparsity at inference time, the model inherently introduces a form of regularization that mitigates overfitting, especially under limited training data or noisy sensor inputs. The ability to deactivate uninformative or redundant modalities also improves robustness under partial modality failures, particularly when vision or lidar becomes unreliable during adverse weather conditions.

In practice, the value of $N$ can be treated as a tunable hyperparameter that balances accuracy and efficiency. A smaller $N$ yields faster inference but may omit useful modality information, while a larger $N$ improves performance but increases resource demands. To enhance stability during training, it is often beneficial to start with a relatively larger $N$ and gradually anneal it using a predefined schedule or entropy-based gating regularization, so that the model progressively learns to rely on a compact subset of experts.
Moreover, the sparse MoE framework is fully compatible with other multimodal extensions, including hierarchical expert routing \cite{li2025hierarchical} and uncertainty-aware gating \cite{zhang2024efficient}. These extensions can further improve specialization and generalization in complex ISAC scenarios, and can be readily integrated on top of the sparse routing mechanism.

In summary, the sparse multimodal MoE framework provides a lightweight yet adaptive solution for efficient ISAC inference. By activating only a subset of experts based on modality-aware gating, the model achieves favorable trade-offs between computational cost, robustness, and prediction performance. In the next section, we evaluate the proposed multimodal MoE and sparse multimodal MoE frameworks under multiple ISAC tasks in realistic LAWN scenarios to quantitatively demonstrate these benefits.

\section{Simulation Results}\label{sec_5}

In this section, we conduct comprehensive simulations to evaluate the effectiveness of the proposed multimodal MoE frameworks in realistic LAWN scenarios. The evaluation covers three representative ISAC tasks under various baseline fusion strategies and performance metrics.

\subsection{Simulation Setups}\label{sec:sim_setup}

\begin{figure}[t]
	\centering
	\vspace{5pt}
	\begin{subfigure}[b]{0.48\textwidth}
		\centering
		\includegraphics[width=\textwidth]{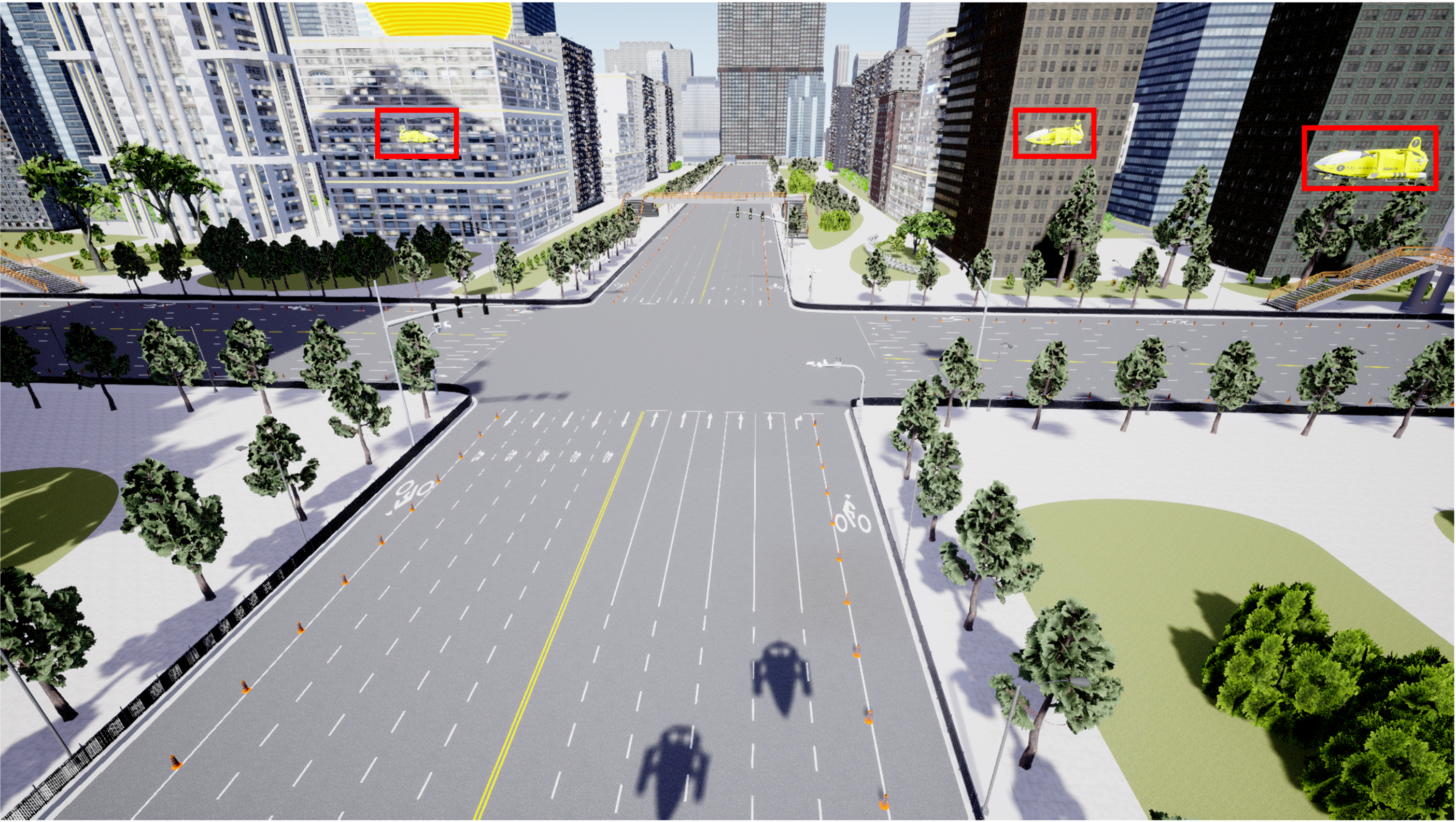}
		\caption{UAV perspective.}
		\label{fig:bs_view}
	\end{subfigure}
	\hfill
	\begin{subfigure}[b]{0.48\textwidth}
		\centering
		\vspace{10pt}
		\includegraphics[width=\textwidth]{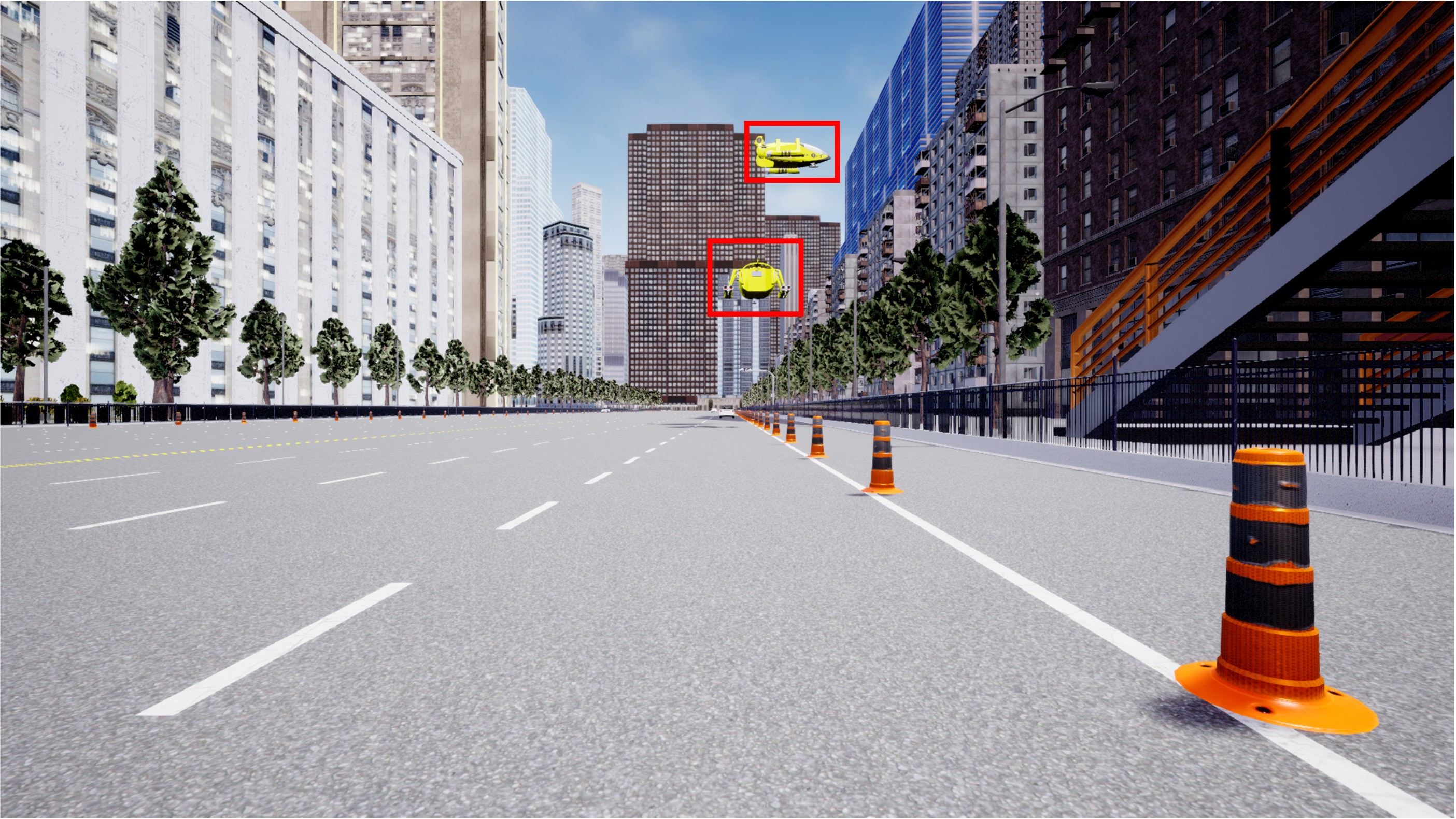}
		\caption{Ground BS perspective.}
		\label{fig:uav_view}
	\end{subfigure}
	\caption{Illustration of a representative scene from the multimodal ISAC dataset in
		LAWNs, showing the environment from the ground BS and UAV perspectives. {UAVs are marked in red boxes}.}
	\label{fig:sim_scene}
\end{figure}

We evaluate the proposed multimodal MoE frameworks using a public multimodal ISAC dataset for low-altitude scenarios~\cite{cheng2025synthsom,10315088}, which provides synchronized RF communication data and heterogeneous sensing modalities.
Fig.~\ref{fig:sim_scene} illustrates a representative scene from the dataset, showing the environment from both the ground BS and UAV perspectives, with UAVs highlighted in red boxes.
We consider three representative ISAC tasks: i) \emph{sensing-aided beam prediction}, where the model predicts the optimal beam index from sensing inputs and is evaluated by top-1 beam accuracy; ii) \emph{sensing-aided path loss prediction}, where the model predicts downlink path loss and is evaluated by the MSE between estimated and ground-truth values; and iii) \emph{communication-aided UAV trajectory tracking}, where the UAV position is regressed from sensing and communication features, and evaluated by MSE and average Euclidean distance between predicted and ground-truth positions.

For the sensing model, each UAV is equipped with an RGB camera and lidar sensor, while the BS carries frequency-modulated continuous wave (FMCW) radars operating at 77–81 GHz. The sensing data includes RGB images, lidar point clouds, mmWave radar point clouds, and GPS position information captured across 3,000 time slots.
For the communication model, we simulate the wireless channels at 5.915 GHz with 20 MHz bandwidth. The UAVs are equipped with 4-element ULAs, and the resulting data includes 30,000 time slots of complex-valued channel impulse responses and path loss values under dynamic propagation conditions. The RF communication and multimodal sensing data are fully synchronized to enable multimodal fusion.

For the learning model, we implement a multimodal MoE architecture with $5$ sensing modalities: vision images, lidar, radar, position information, and RF communication signals. Each modality is assigned a group of 3 parallel experts, resulting in a total of 15 expert subnetworks.
In the sparse MoE variant, we activate a subset of 5 experts out of the total 15 for each input, and their weights are renormalized prior to aggregation.
For image-based modalities, each expert adopts a ResNet-18 backbone \cite{he2016deep}; for point-cloud-based modalities, each expert is implemented using a PointNet backbone \cite{qi2017pointnet}. The modality-aware gating network is a three-layer MLP with 128 hidden units and ReLU activations, followed by a softmax layer to produce normalized fusion weights.
All models are trained using the Adam optimizer with a learning rate of $10^{-3}$ and batch size of $64$. We split the entire dataset into 70\% for training and 30\% for testing, and all results are reported on the testing set.

To evaluate the effectiveness of the proposed multimodal MoE and sparse MoE frameworks, we compare them against the following representative baselines.

\begin{itemize}[leftmargin=1.5em]
	\item \textbf{Single-modality baselines} \cite{9129369, 9939167}: We include vision-only and lidar-only models, where each modality is processed independently using the same expert network architecture without any fusion. These serve as lower-bound references for unimodal performance.
	
	\item \textbf{Feature concatenation fusion} \cite{10912462}: Multimodal features from all modalities are concatenated after backbone encoding and fed into a shared MLP prediction head. This represents a common and straightforward fusion strategy.
	
	\item \textbf{Multimodal transformer} \cite{10539181, 11208675}: A transformer encoder is employed to model cross-modal interactions by attending over modality-specific token embeddings. Self-attention and cross-attention mechanisms enable the network to dynamically capture inter-modal dependencies and contextual correlations across sensing streams.
	
	\item \textbf{Multimodal LLM} \cite{10829757}: We adopt a pretrained LLM by projecting sensor features into a shared embedding space followed by fine-tuning. This baseline highlights the performance of large-scale pretrained fusion.
\end{itemize}

\begin{figure}[t]
	\renewcommand\figurename{Fig.}
	\centering \vspace*{5pt} \setlength{\baselineskip}{10pt}
	\includegraphics[width = 0.49\textwidth,trim=10 10 10 0,clip]{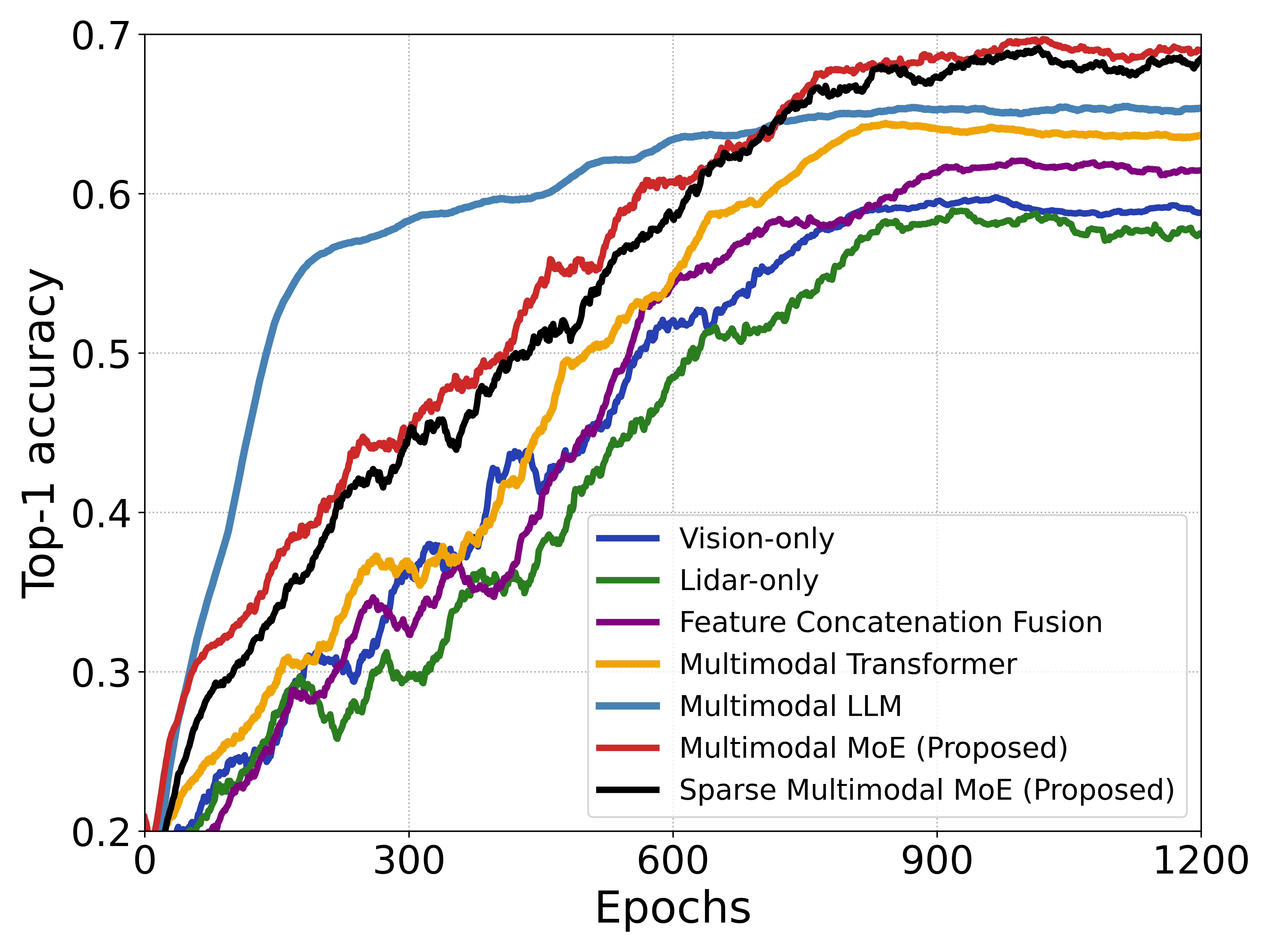}
	\caption{Comparison of Top-1 beam prediction accuracy for the proposed multimodal MoE frameworks and benchmark methods.}
	\label{fig:beam_prediction}
\end{figure}

\subsection{Sensing-Aided Beam Prediction}

In this subsection, we evaluate model performance on the sensing-aided beam prediction task, where the goal is to select the optimal downlink beam index based on multimodal sensing inputs. As shown in Fig.~\ref{fig:beam_prediction}, the proposed multimodal MoE and sparse MoE frameworks significantly outperform all baselines in terms of top-1 prediction accuracy. The dense multimodal MoE model achieves the best final performance, while the sparse MoE maintains competitive accuracy with fewer active experts, demonstrating its efficiency. The superior results can be attributed to the adaptive gating mechanism, which assigns dynamic fusion weights to expert subnetworks, allowing the model to selectively emphasize informative modalities under varying channel and environmental conditions.

Among the baselines, the multimodal transformer and multimodal LLM benefit from stronger representation capacity and attention mechanisms, achieving better performance than early fusion via feature concatenation. However, their performance remains below the MoE-based models due to the lack of explicit expert specialization across modalities. Unimodal baselines (vision-only and lidar-only) show limited accuracy, confirming that leveraging diverse and complementary sensing modalities is essential for robust beam prediction in dynamic low-altitude environments.

\begin{figure}[t]
	\renewcommand\figurename{Fig.}
	\centering \vspace*{5pt} \setlength{\baselineskip}{10pt}
	\includegraphics[width = 0.49\textwidth,trim=10 10 10 0,clip]{figs/sim2.png}
	\caption{Comparison of NMSE performance for sensing-aided path loss prediction using the proposed multimodal MoE frameworks and benchmark methods.}\label{fig:path_loss}
\end{figure}

\subsection{Sensing-Aided Path Loss Prediction}

In this subsection, we evaluate all methods on the task of sensing-aided path loss prediction, where the goal is to estimate the downlink path loss using multimodal sensing features. Fig.~\ref{fig:path_loss} reports the normalized mean squared error (NMSE) curves in dB scale over training epochs. The proposed multimodal MoE achieves the lowest NMSE, followed closely by the sparse MoE, which performs nearly on par while requiring fewer active experts. These results highlight the advantage of using adaptive modality-aware expert selection to adaptively predict path loss under varying environmental conditions and blockage states.

Compared to the baselines, the multimodal transformer and LLM deliver improved performance over early fusion approaches by modeling cross-modal correlations and utilizing large-scale pretraining, respectively. However, they remain outperformed by MoE-based methods due to their lack of explicit expert specialization and adaptive routing. Meanwhile, unimodal baselines (vision-only and lidar-only) suffer from limited expressiveness, reflecting the insufficiency of single-modality inputs in accurately capturing fine-grained propagation loss variations in complex environments.

\subsection{Communication-Aided UAV Trajectory Tracking}

In this subsection, we evaluate the proposed multimodal MoE method and sparse MoE on the task of UAV trajectory tracking, where the objective is to predict the UAV's 3D position at each time step based on multimodal sensing and RF communication inputs. This task is essential for autonomous navigation in dynamic low-altitude environments, where accurate trajectory estimation enables proactive control and reliable mobility support.

\begin{figure}[t]
	\renewcommand\figurename{Fig.}
	\centering \vspace*{5pt} \setlength{\baselineskip}{10pt}
	\includegraphics[width = 0.49\textwidth,trim=10 10 10 0,clip]{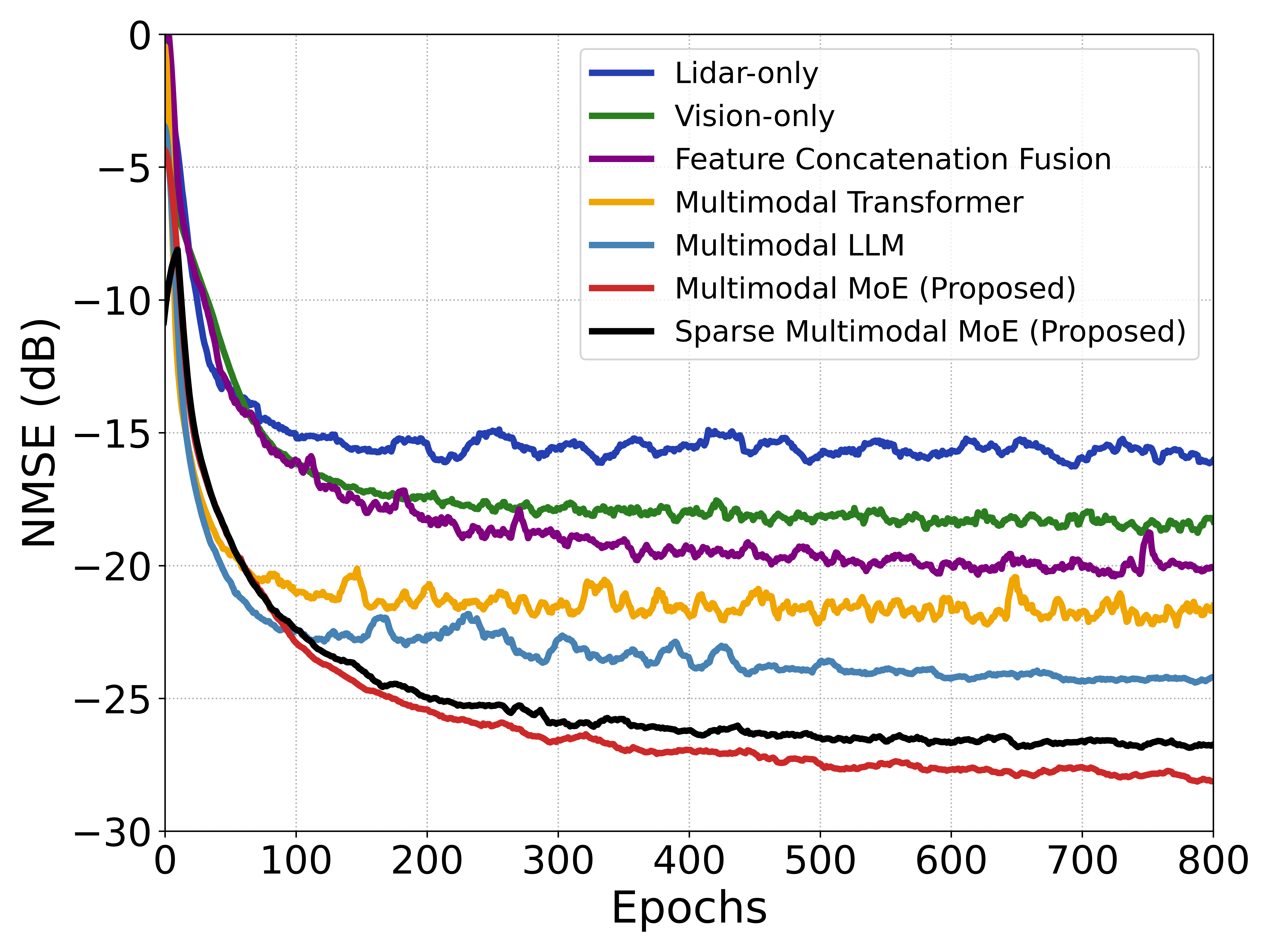}
	\caption{Comparison of NMSE performance for communication-aided UAV trajectory tracking using the multimodal MoE frameworks and benchmark methods.}\label{fig:sim3}
\end{figure}

\begin{figure}[t]
	\renewcommand\figurename{Fig.}
	\centering
	\setlength{\baselineskip}{10pt}
	\begin{subfigure}{0.48\textwidth}
		\centering
		\includegraphics[width=\linewidth,trim=50 50 60 60,clip]{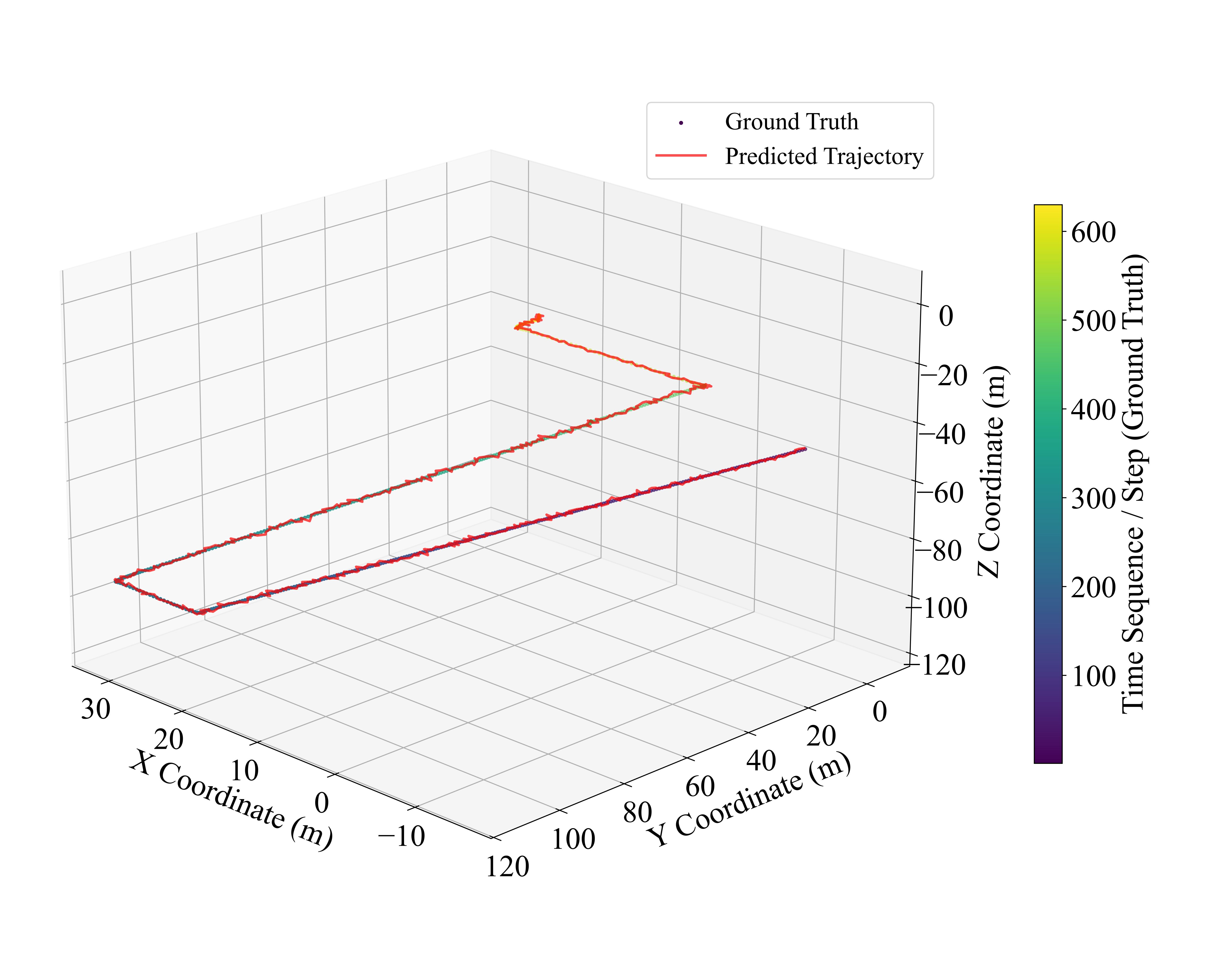}
		\caption{Scenario 1}
		\label{fig:tra1}
	\end{subfigure}
	\vspace{5pt}
	
	\begin{subfigure}{0.48\textwidth}
		\centering
		\includegraphics[width=\linewidth,trim=50 50 60 60,clip]{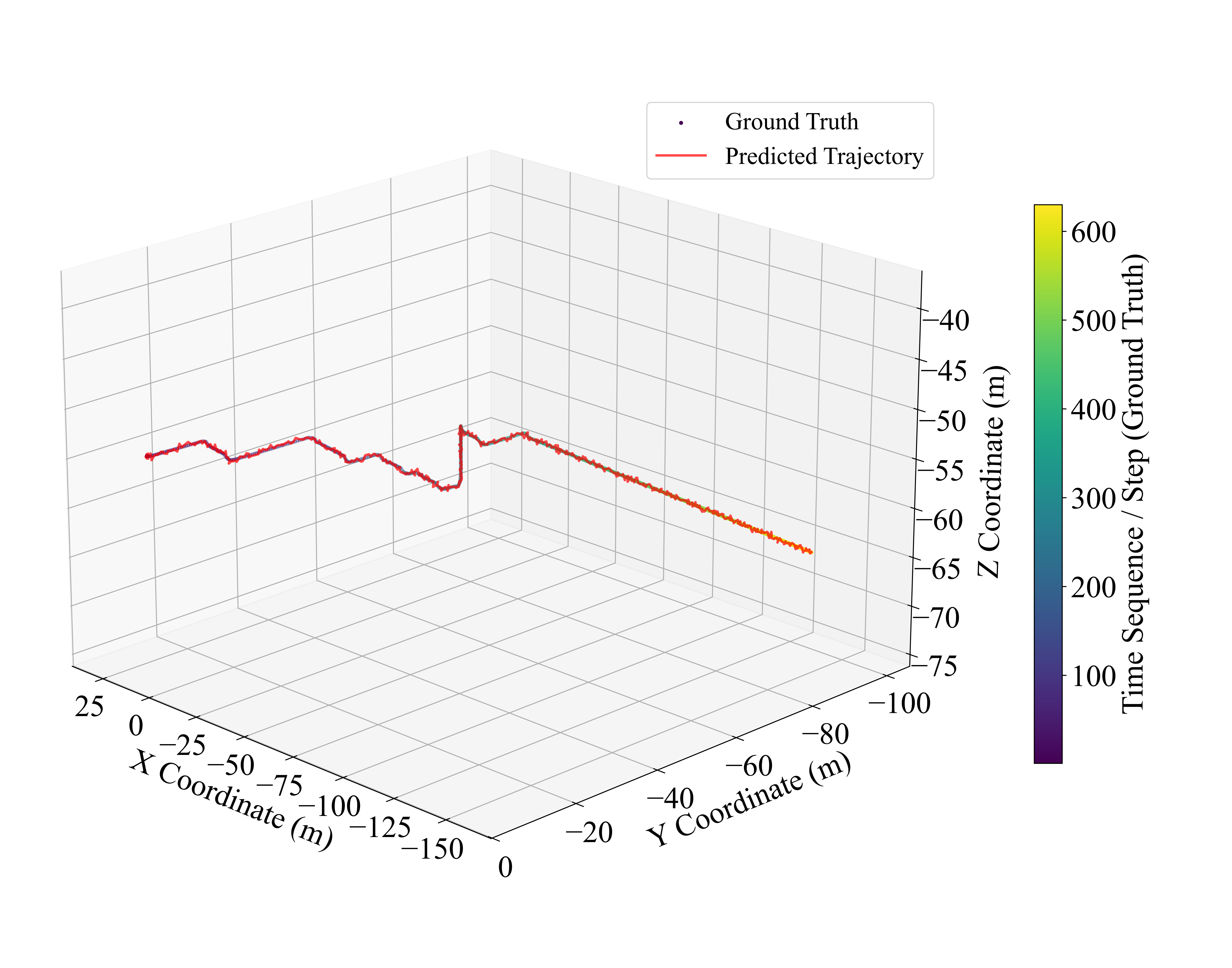}
		\caption{Scenario 2}
		\label{fig:tra2}
	\end{subfigure}
	
	\caption{Comparison between predicted and ground-truth UAV trajectories using the proposed multimodal MoE framework.}
	\label{fig:uav_trajectory}
\end{figure}

Fig.~\ref{fig:sim3} shows the NMSE in dB scale across training epochs. As observed, the proposed multimodal MoE achieves the lowest NMSE, indicating superior estimation accuracy, while the sparse MoE maintains comparable performance with reduced computational overhead. These results validate the effectiveness of adaptive expert selection in capturing complex path-dependent signal variations. Among the baselines, the multimodal transformer and multimodal LLM achieve moderate improvements over the static fusion baseline by leveraging attention mechanisms and pretrained multimodal embeddings, respectively. However, their fusion lacks explicit modality specialization, resulting in inferior performance compared to the MoE-based architectures. Unimodal methods (vision-only and lidar-only) consistently underperform, highlighting the limitations of single-modality inputs in accurately reconstructing continuous trajectories amidst sensor drift and environmental blockage.

To further illustrate the model’s performance, Fig.~\ref{fig:tra1} and Fig.~\ref{fig:tra2} visualize the predicted trajectory generated by the multimodal MoE alongside the ground-truth UAV path. The predicted positions exhibit close alignment with the true trajectory, particularly around direction changes and elevation variations, demonstrating the model’s ability to accurately track UAV motion in dynamic low-altitude scenarios.

\section{Conclusion}

In this paper, we have investigated multimodal ISAC in LAWNs, leveraging heterogeneous sensing modalities and communication signals to achieve robust environmental perception and reliable connectivity in dynamic aerial environments.
We have formulated a general learning-based optimization framework for multimodal ISAC, which has been instantiated across three representative ISAC tasks. To address the challenges of modality heterogeneity and time-varying reliability, we have proposed a multimodal MoE architecture that adaptively assigns modality-specific fusion weights via a learnable gating network. Additionally, we have developed a sparse MoE variant that reduces inference costs by selectively activating a subset of expert networks, while preserving the benefits of adaptive fusion.
Extensive simulations on three representative ISAC tasks in LAWNs have demonstrated that the proposed methods consistently outperform conventional multimodal fusion baselines in terms of learning performance and training sample efficiency.

\bibliographystyle{ieeetr}
\bibliography{IEEEabrv,refs}


\end{document}